\documentclass[preprint,proceedings,nonatbib]{rmaa}
\usepackage{OLDLATEX/natbib}

\suppressfulladdresses % by popular request

\usepackage{paralist}

\usepackage{psfrag,color}

\def\ltsima{$\; \buildrel < \over \sim \;$}
\def\simlt{\lower.5ex\hbox{\ltsima}}            % < over ~
\def\gtsima{$\; \buildrel > \over \sim \;$}
\def\simgt{\lower.5ex\hbox{\gtsima}}            % > over ~

\def\sun{\odot}
\def\ms{\ensuremath{\mathrm{\,M_{\sun}}}}
\def\ergsec{erg\,s$^{-1}$}
\def\grcm3{g\,cm$^{-3}$}

\SetYear{2006} \SetConfTitle{Circumstellar Media and Late Stages of
  Massive Stellar Evolution}

\title{Relativistic Outflows in Gamma-Ray Bursts} 

\author{M. A. Aloy\altaffilmark{1,2}, and 
        M. Obergaulinger\altaffilmark{2}}

\altaffiltext{1}{Departamento de Astronom\'\i{}a y Astrof\'\i{}sica,
  Universidad de Valencia, 46100, Burjassot, Espa\~na (Miguel.A.Aloy@uv.es).}

\altaffiltext{2}{Max-Planck-Institut f\"ur Astrophysik,
  Karl-Schwarschild-Str. 1, 85741 Garching, Germany}

\shortauthor{Aloy \& Obergaulinger}
\shorttitle{Relativistic outflows in GRBs}

\listofauthors{M. A. Aloy, \& M. Obergaulinger}
\indexauthor{Aloy, M. A.}
\indexauthor{Obergaulinger, M}

\abstract{The possibility that gamma-ray bursts (GRBs) were not
  isotropic emissions was devised theoretically as a way to ameliorate
  the huge energetic budget implied by the standard fireball model for
  these powerful phenomena.  However, the mechanism by which after the
  quasy-isotropic release of a few $10^{50}\,$erg yields a collimated
  ejection of plasma could not be satisfactory explained
  analytically. The reason being that the collimation of an outflow
  by its progenitor system depends on a very complex and
  non-linear dynamics. That has made necessary the use of numerical
  simulations in order to shed some light on the viability of some
  likely progenitors of GRBs. In this contribution I will review the
  most relevant features shown by these numerical simulations and how
  they have been used to validate the collapsar model (for long GRBs)
  and the model involving the merger of compact binaries (for short
  GRBs).}

\resumen{La posibilidad de que las erupciones de rayos gamma (GRBs) no
  sean emisiones isotr\'opicas fue considerada te\'oricamente para
  atenuar el problema asociado a la gran cantidad de energ\'{\i}a
  implicada por el modelo de bola de fuego est\'andar para estos
  potentes fen\'omenos. Sin embargo, el mecanismo por el cual, tras la
  deposi\'on cuasi-isotr\'opica de unos pocos $10^{50}\,$erg se
  origina una eyecci\'on colimada de plasma no pudo ser explicada de
  forma satisfactoria anal\'{\i}ticamente. La raz\'on de ello radica
  en que la colimaci\'on de un flujo saliente por su sistema
  progenitor depende de una din\'amica no lineal muy compleja. Ello
  ha hecho necesario el uso de simulaciones num\'ericas para arrojar algo
  de luz sobre la viabilidad de algunos de los progenitores m\'as
  probables de GRBs. En esta contribuci\'on revisar\'e los hechos
  m\'as relevantes mostrados por tales simulaciones num\'ericas y
  c\'omo \'estas han sido utilizadas para validar el modelos de
  estrella colapsante (para GRBs largos) y el modelo que implica la
  fusi\'on de un sistema binario de objetos compactos (para GRBs
  cortos).}

\addkeyword{gamma-rays: bursts}
\addkeyword{ISM: Jets and outflows}
\addkeyword{MHD}
\begin{document}
% Typeset article header
\maketitle

\section{Introduction}
\label{sec:intro}

Our current understanding is that gamma-ray bursts (GRBs) are produced
in the course of the birth of a stellar-mass black hole (BH). In other
astrophysical systems where accreting BHs fuel collimated beams of
plasma (e.g., AGNs and BH--X-ray binaries), there is a direct evidence
for relativistic outflows and jet collimation which comes from the
imaging of the system. Therefore, it seems reasonable to assume as
starting point, that also GRBs result from relativistic, collimated
outflows from accreting, stellar-mass BHs.  We have inferred that
outflows yielding GRBs are relativistic because of a couple of
observational constraints, namely, the detection of radio
scintillation of the interstellar medium \citep{FK97} and the
measurement of superluminal proper motions in imaged afterglows
\citep{TFBK04}. The ultrarelativistic expansion is also necessary to
overcome the theoretical constrain imposed by the compactness problem
\citep{CR78}. However, we only have an indirect evidence of
collimation based on the observational constraint posed by the
achromatic break in the afterglow light curve of some GRBs (e.g.,
\citealp{Harrisonetal99}). From the theoretical point of view, if GRBs
are collimated events, the true emitted energy $E_\gamma$ is reduced
by a factor $f_{\Omega} \simeq \theta^2/2$ \citep{Rhoads99,SPH99},
i.e., $E_\gamma = f_\Omega E_{\gamma, {\rm iso}}$, where $ E_{\gamma,
  {\rm iso}}$ is the detected equivalent isotropic
energy. Nonetheless, the mechanism by which after the quasi-isotropic
release of an amount of energy in the range $10^{48} - 10^{51}\,$erg
results in a collimated ejection has not satisfactory been explained
analytically. The reason being that the collimation of an outflow by
the progenitor system depends on a very complex and non-linear
dynamics. That has made necessary the use of numerical simulations in
order to understand the collimation mechanism as well as to shed some
light on the viability of some systems proposed to be the progenitors
of GRBs.  A robust result obtained in numerical simulations of
generation of GRBs is that the progenitor system yields collimated
outflows under rather general conditions independent on whether the
outflow is initiated thermally (e.g., \citealp{Alo00b,AJM05})
or it is magnetically driven \citep{McKinney06}.

In this contribution I will review the most relevant features shown by
these numerical simulations and how they have been used to validate
the collapsar model (for long GRBs; \S~\ref{sec:progen_lGRBs}) and the
model involving the merger of compact binaries (for short GRBs;
\S~\ref{sec:progen_sGRBs}). 

\section{Outflows emerging from progenitors of long GRBs}
\label{sec:progen_lGRBs}

Among the plethora of models devised to explain the origin of long
GRBs (lGRBs), the most widely accepted was put forward by Woosley. In
the original {\it collapsar} model, also known as {\it failed
  supernova} model \citep{Wo93}, the collapse of a massive
($M_\mathrm{ZAMS} \sim 30\ms$) rotating star that does not form a
successful supernova but collapses to a BH ($M_\mathrm{BH}\sim 3\ms$)
surrounded by a thick disk. The viscous accretion of the disk matter
onto the BH yields a strong heating that, in its turn, produces a
copious amount of thermal neutrinos and antineutrinos, which
annihilate preferentially around the rotation axis producing a
fireball of $e^+e^-$pairs and high energy photons. Later it was noted
that, perhaps $\nu$--powered fireballs might not be sufficiently
energetic to fuel the most powerful GRB events and, thus, the
collapsar model was extended to account for alternative energy
extraction mechanisms \citep{MWH01}.  More explicitely, the accretion
energy of the torus could be tap by sufficiently strong magnetic
fields ({\em hydromagnetic} generation) by means of the
Blandford-Payne process \citep{BP82}, or a non-trivial fraction of the
rotational energy of the BH may also be converted into a Poynting flux
\citep{BZ77}.

The scape of the newly born fireball and its terminal Lorentz factor
($\Gamma_\infty$) depend on structural and on dynamical factors. The
critical structural factors are the environmental baryon density in
the funnel around the rotation axis of the star and the ability of the
progenitor star to loose its outer Hydrogen envelope. An under-dense
funnel forms along with the accretion torus if the specific angular
momentum of the core of the star lies in the range
$3\times10^{16}\,$cm$^2\,$s$^{-1} \simlt j \simlt 2 \times
10^{17}\,$cm$^2\,$s$^{-1}$ \citep{MW99}. The existence of the funnel
is key to collimate the fireball and to permit its propagation through
the progenitor. The most favourable conditions for the generation and
propagation of the fireball happen when the density of the funnel
($\rho_\mathrm{f}$) is much smaller than the density of the torus
($\rho_\mathrm{torus}$), namely, $\rho_\mathrm{f}/\rho_\mathrm{torus}
\simlt 10^{-4}-10^{-3}$ \citep{MW99}. The likelihood that the
progenitor star had lost its Hydrogen envelope depends on a number of
factors which occurrence is still a matter of debate like, e.g., the
generation of stellar winds, the interaction with a companion
\citep{Podsialdlowskietal04}, etc. The relevance of the lost of the
hydrogen envelope resides on the fact that, unless the density of the
funnel is extremely small (as proposed by \citealp{MR01}), only a
mildly relativistic, poorly-collimated fireball would reach the outer
edge of the hydrogen envelop after very long times and with relatively
small Lorentz factors ($\Gamma \simeq 2$) implying that the
observational signature would be an X--ray/UV transient with a
duration of $\sim 100 - 1000\,$s but not a GRB \citep{MWH01}. Finally,
the most important dynamical factor setting $\Gamma_\infty$ is the the
amount of baryons entrained as the fireball propagates through the
stellar mantle.

Several analytic works have made estimates about the collimation angle
and the Lorentz factor when the fireball breaks out the surface of the
star (e.g., \citealp{MR01}). Nevertheless, the complexity inherent to
the non-linear (magneto)hydrodynamic interaction of the fireball
plasma with the stellar environment makes it unavoidable the use of
numerical simulations. With these simulations we have been able to
give preliminary answers to the following questions:

\noindent {\bf Collimation}.  The generated outflows are inertially or
magnetically confined. In the first case, the collimator is the funnel
within the progenitor while in the second case, the flow is
self-collimated by its own magnetic field if it is strong
enough. Rather independently on the initial conditions and on the
inclusion of magnetic fields, the typical outflow half-opening angles,
when the jet reaches the surface of the progenitor star, are
$\theta_{\rm break} \simlt 5^o$. These small half-opening angles
result from the recollimation of the outflow within the progenitor and
they are independent on whether the boundary conditions are set to
initiate the jet with much larger half-opening angles (e.g.,
$\theta_0=20^o$; \citealp{ZWM03}) or on whether the jet is generated
by an energy release into a volume spanning a half-opening angle
$\theta_\mathrm{d}=30^o$ much larger than $\theta_{\rm break}$
\citep{Alo00b}. In the course of their propagation through the
progenitor the jets develop a non-homogeneous structure, transverse to
the direction of motion, whose main features are an internal
ultrarelativistic spine (where the Lorentz factor may reach
$\Gamma_\mathrm{core} \sim 30-50$ at jet breakout) within a
half-opening angle of $<2^o$ laterally endowed by a moderately
relativistic, hot shear layer ($\Gamma_\mathrm{shl} \sim 5-10$)
extending up to $\theta_\mathrm{shl}<20^o-30^o$ \citep{Alo02}. We
point out that the transverse structure of the jet is {\it nearly}
Gaussian both in simulations including magnetic fields (e.g., McKinney
2006) or not including them (e.g., \citealp{Alo00b}). However, a more
accurate fit of the transverse structure of the jet, cannot be
accommodated by a simple Gauss function \citep{Aloy01}.

\noindent
{\bf Variability}. All produced outflows are highly variable due to
the generation of Kelvin-Helmholtz \citep{Alo00b,GH04}, shear-driven
\citep{Alo02} or pinch magnetohydrodynamic (MHD) instabilities
\citep{McKinney06}. Such {\it extrinsic} variability is independent on
the ({\it intrinsic}) variability of the energy source and leads to
the formation of irregularities in the flow which are the seeds of
internal shocks in the outflow. Except in cases in which the source
may produce quasi-periodic variability (perhaps induced by precession
or nutation modes of the accretion disc), the extrinsic variability
might be indistinguishable form the intrinsic one. Numerical
simulations of three-dimensional (3D) relativistic jets propagating
through collapsar-like environments show that such jets are also
stable \citep{ZWH04} but it still remains unknown whether 3D
relativistic, magnetohydrodynamic, collapsar-jets will also be stable
along its whole trajectory.

\begin{figure*}[!t]
\begin{tabular}{cc}
  \includegraphics[width=0.92\columnwidth]{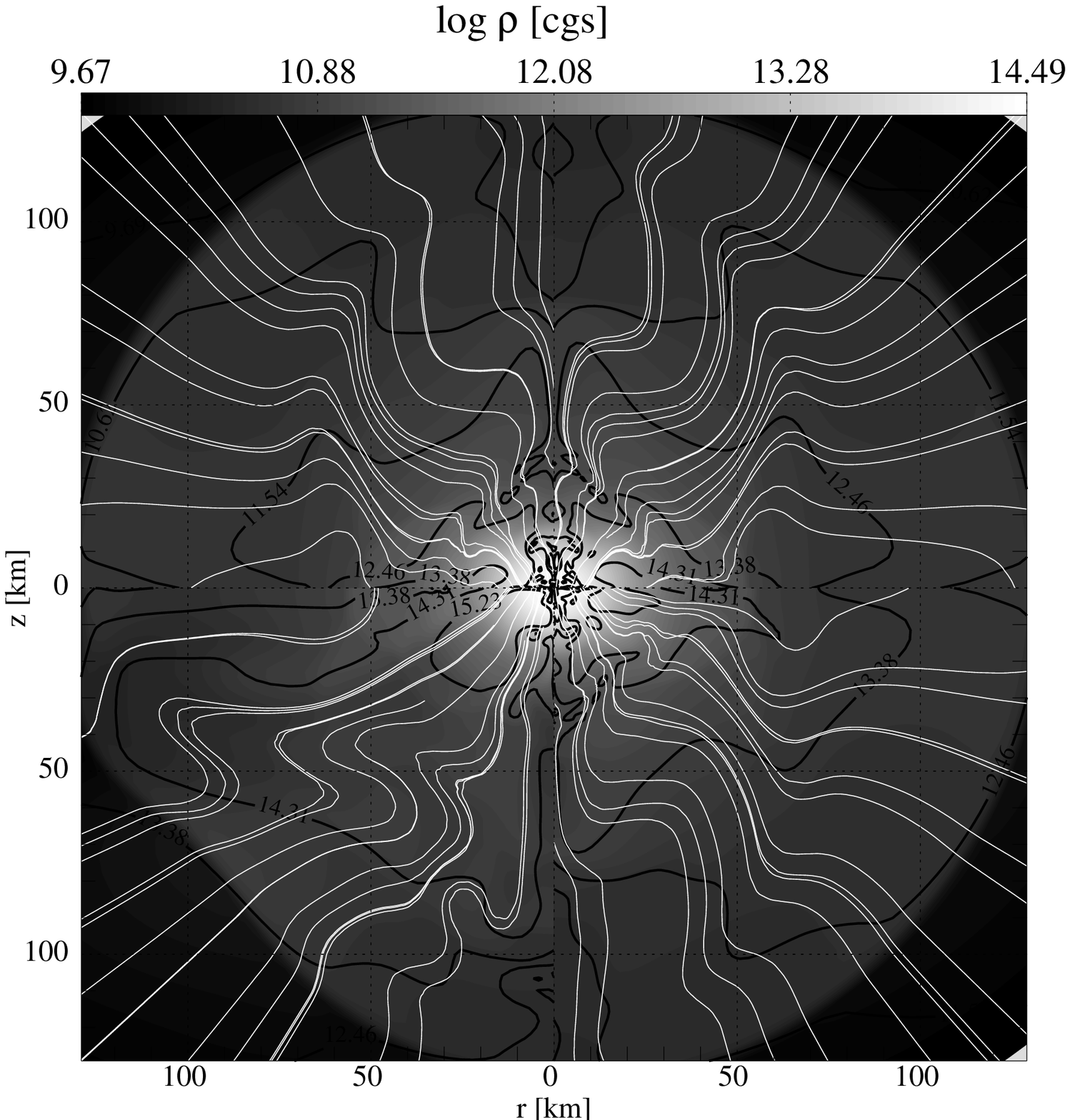} &
  \includegraphics[width=0.92\columnwidth]{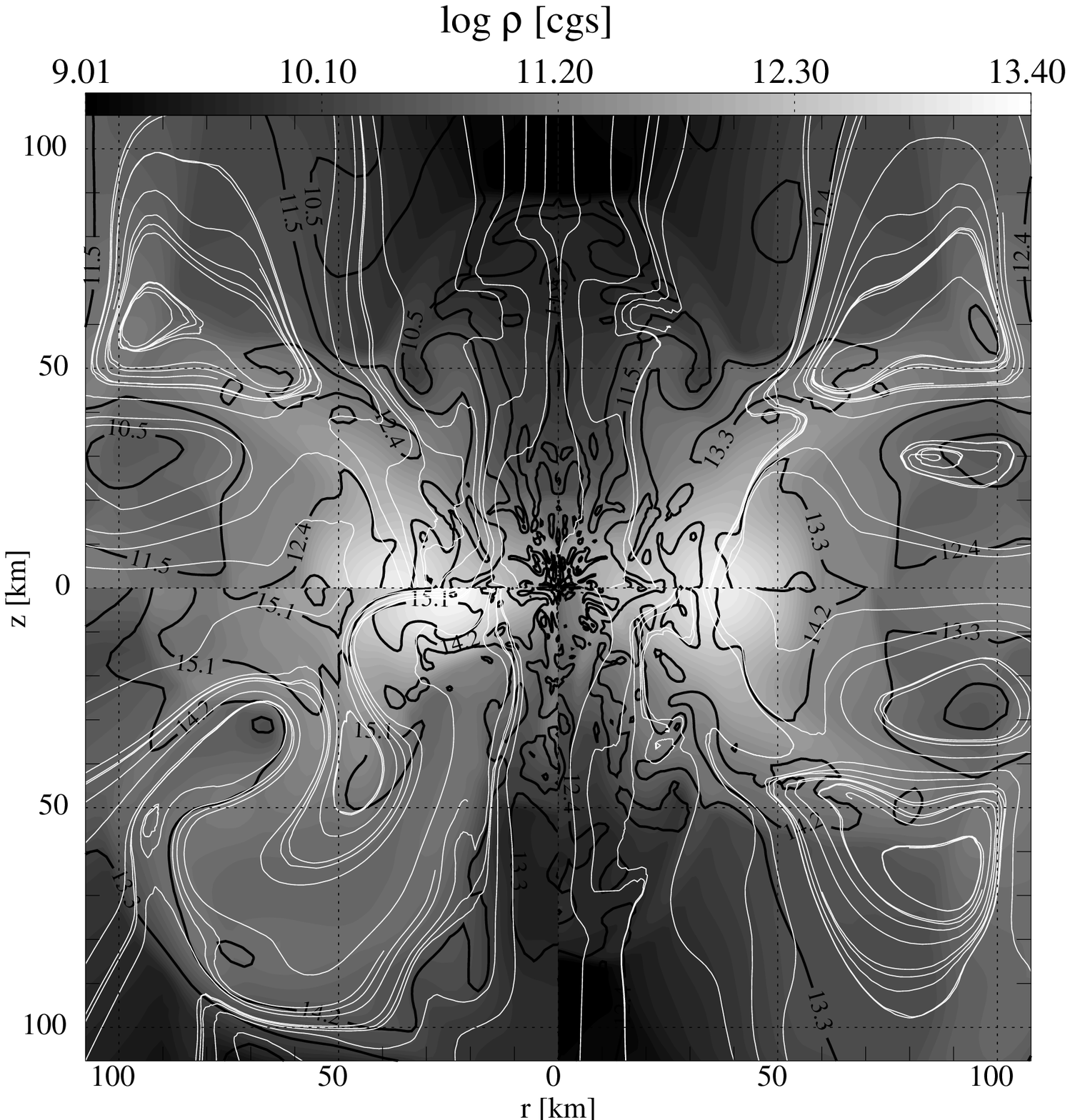} 
\end{tabular}
\caption{Logarithm of the density (gray-scale) with overimposed
  magnetic field lines (white lines) and total magnetic field strength
  (black contours) corresponding to models collapsing stellar cores
  with a small initial rotational energy and rotating almost rigidly
  (left) and with a larger initial rotational energy and
  differentially rotating (right). Each panel displays the state of
  the stellar core after the bounce for four different initial
  magnetic field strengths ($B_0=10^{10}\,$G, $10^{11}\,$G,
  $10^{12}\,$G and $10^{13}\,$G in the clockwise direction starting
  from the upper left corner). The figures correspond to models of
  Obergaulinger et~al.\@ (2006a).}
  \label{fig:initialB}
\end{figure*}

\noindent {\bf Jet breakout}. The jets generated are much lighter than
their baryon reach environments. Thus, they propagate through the
collapsar at moderate speeds ($\sim 0.3c$) and fill up thick
cigar-shaped cavities or cocoons of shocked matter that also
propagates along with the beam of the jet and that, eventually may
break the surface of the collapsar. Since in the cocoon a few
$10^{50}\,$erg may be stored as the jet drills its way through the
star, it has been proposed that its eruption through the collapsar
surface could yield a number of $\gamma$-ray/X-ray/UV-transients
\citep{RRCR02}.  Indeed, it has been proposed that GRBs are but one
observable phenomenon accompanying black hole birth and other
possibilities may arise depending on the observer's viewing angle with
respect to the propagation of the ultrarelativistic jet
\citep{Woosley00}. Thus, in a sort of unification theory for
high-energy transients, one may see progressively softer events
ranging from GRBs (when the jet emergence is seen almost head on) to
UV flashes (when the jet eruption is seen at relatively large polar
angles) and accounting for X-ray rich GRBs (XRR-GRBs) and X-ray
flashes (XRF) at intermediate angles. The jet emergence through the
stellar surface and its interaction with the stellar wind (which
likely happens during the late stages of the evolution of massive
stars) could lead to some precursor activity \citep{MWH01}.
Furthermore, $\nu$--powered jets are very hot at breakout ($\sim 80\%$
of the total energy is stored in the form of thermal energy) which
implies that jets can still experience an additional acceleration by
conversion of thermal into kinetic energy, even if the energy source
has ceased its activity.

\subsection{lGRBs produced in collapsars: MHD- or neutrino-powered jets?}

From a aesthetic point of view, it is beautiful if any invoked jet
powering mechanism explains not only events with relatively {\it
  small} Lorentz factors $\Gamma \sim 100$ but also the occurrence of
events with very large inferred values of $\Gamma \sim 500$
\citep{LS01} or even $\Gamma \sim 1000$ as suggested by models for
some GRBs \citep{SRR03}. However, there are no fundamental reasons to
argue in favour of a unique and universal mechanism to extract the
energy stored in the progenitor.

Purely hydrodynamic, $\nu$-powered jets in collapsars of type-I seem
to be able only to produce more moderate terminal values of the bulk
Lorentz factor ($\Gamma \sim 100-400$; e.g., \citealp{Alo00b,ZWM03}),
even if there is a further acceleration of the forward shock as a
result of an appropriate density gradient in the medium surrounding
the progenitor \citep{Alo00b}, unless the density in the funnel around
the rotation axis is very small \citep{MR01}. In collapsars of type-II
the mass accretion rate ($\dot{M} \sim 10^{-3}-10^{-2}\ms\,$s$^{-1}$)
is insufficient to produce a $\nu$-powered jet, but it may suffice to
generate MHD-powered jets whose observational signature might be weak,
poorly collimated X-ray/UV transients or very long ($\sim 100 -
1000\,$s) GRBs of low $E_{\rm iso}$ and small $\Gamma$ if the
progenitor star is able to loose its Hydrogen envelope \citep{MWH01}.

From extrapolations of the numerical results of axisymmetric jets
generated electromagnetically \citep{McKinney06} very large values of
the asymptotic Lorentz factor of the outflow ($\Gamma_\infty \sim
1000$) can be attained. Thus, it seems rather plausible that MHD
mechanisms have to be employed to generate jets with $\Gamma \sim
500-1000$ \citep{McKinney05,McKinney06} and, invoking the existence of
a universal energy extraction mechanism, one may also argue that also
jets with much more moderate terminal Lorentz factors are also
MHD-generated. Nevertheless, these arguments have a number of issues:

{\bf 1.-} The estimates of the terminal Lorentz factor of
MHD-generated jets are based on axisymmetric models. The 3D-stability
of relativistic magnetized jets is still a matter of debate which
needs of much more complex numerical simulations than the ones
produced so far. Should 3D-MHD jets be unstable, should the terminal
Lorentz factor not reliably be calculated with the state-of-the-art
axisymmetric, general relativistic, MHD (GRMHD) simulations up to date
\citep{McKinney06}.

{\bf 2.-} GRMHD initial models of accreting BH systems consist on more
or less realistic matter distributions over which an assumed poloidal
field is imposed, i.e., they are not the result of a consistent
magneto-rotational core collapse (e.g.,
\citealp{Mizuetal04a,Mizuetal04b,McKinney06}).  The use of poloidal
fields is triggered by the fact that purely toroidal field
configurations do not yield the production of bipolar jets
\citep{Villiersetal05}. On the other hand, the initial magnetic field
strengths are assumed, not consistently computed from the core
collapse of massive stars. Typically, initial field strengths as large
as $B \sim 10^{15} - 10^{16}\,$G are used. Such large values of $B$ in
combination with maximal values of the dimensionless angular momentum
of the BH ($j_\mathrm{BH} \sim 1$) are necessary in order to
efficiently extract energy via Blandford-Znajek (BZ) mechanism,
because the BZ-power scales rather sensitive with $B$ as
$\dot{E}_\mathrm{BZ} \sim 10^{50} j_\mathrm{BH}^2 (M_\mathrm{BH} /
3\ms)^2 (B / 10^{15}\,\mathrm{G})^2\,$erg\,s$^{-1}$, \citep{LWB00}.

{\bf 3.-} When numerical simulations of the magnetorotational core
collapse of massive stars are performed (see, e.g.,
\citealp{OAM06,OADM06}, and references therein), bipolar outflows as
well as thick toroidal structures surrounding a central, low density
funnel in the collapsed core are only generated for initial magnetic
field strengths $B_0 \simgt 10^{12}\,$G and for rather
fast\footnote{The rotational energy being as large as $\sim 4\%$ of
  the gravitational energy.} and differentially rotating stellar cores
(Fig.~\ref{fig:initialB} right). Such initial field strengths and
angular velocities are well beyond the ones predicted by the
state-of-the-art calculations of rotating massive stars
\citep{HWS05}. These initially strongly magnetized models likely
develop collapsed cores, with a mass of $M_\mathrm{c} \sim 1\ms$ and a
specific angular momentum $J_\mathrm{c} \sim
10^{16}\,$cm$^2\,$s$^{-1}$, and may form a rapidly rotating BH with
$j_\mathrm{BH} \sim 1$. Furthermore, the winding up of the initial
poloidal field leads to maximal field strengths $B_\mathrm{max} \sim
10^{15}\,$G, being the field predominantly toroidal
$B_\mathrm{toroidal}/B_\mathrm{poloidal}\sim 1- 10$. On the other
hand, initially rigid, more moderately rotating cores do not yield
tori around low-density funnels, do not produce bipolar jets and the
maximum field strengths are $\sim 5\times 10^{14}\,$G
(Fig.~\ref{fig:initialB} left). Considering that the collapsed cores
are smaller ($M_\mathrm{c} \sim 0.75\ms$) and slowly rotating
($J_\mathrm{c} \sim 2\times 10^{15}\,$cm$^2\,$s$^{-1}$) than in the
previous case one may expect that the newly born BH resulting from the
posterior evolution of these kind of cores will not be a maximally
rotating but, instead, they will form BHs with more moderate
$j_\mathrm{BH} \sim 0.6$.  Thereby, the conclusion that seems to
emerge from detailed numerical simulations of Obergaulinger et al.\@
is that if the initial magnetic field strength and rotational energy
is as small as predicted by the most detailed stellar evolution
models, the collapsed core does not hold the appropriate conditions
($B \simgt 10^{15}\,$G, $j_\mathrm{BH} \sim 1$) to efficiently extract
energy via BZ-mechanism and, conversely, {\it unrealistically} large
initial magnetic fields and rotational energies need to be invoked to
expect a BZ-like mechanism to operate efficiently. However, we have to
be cautious in order not to extract too far reaching conclusions from
the previously mentioned numerical work. It remains true that the
results of the simulations of Obergaulinger et al.\@ are handicapped
because they do not include general relativity, because the numerical
resolution is still small to capture all the relevant
magneto-rotationally unstable modes and, because they are restricted
to axisymmetric models.

From the above points, one may infer that the dynamical relevance of
the magnetic field in the process of energy extraction from the
central source will depend on fine details of the magnetorotational
collapse of the collapsar core. On the other hand, the process of
$\nu\bar\nu$-annihilation as the primary source of energy that fuels
an ultrarelativistic fireball also needs of a more careful study in
order to know how much energy such a process may release in the
progenitor system and how such an amount of energy depends on the
physical conditions of the progenitor.

A step towards such goal is the work of Birkl et al.\@ (2006), which
contributes to better understand how the energy deposition rate due to
the process of $\nu\bar\nu$-annihilation ($\dot{E}_{\nu\bar\nu}$)
depends on general relativistic (GR) effects and on different
neutrinosphere geometries in hyperaccreting stellar-mass BH
systems. Birkl et al.\@ consider two families of neutrinospheres. On
the one side, idealized geometries as thin disks, tori, and
spheres. On the other side, more realistic models are constructed as
non-selfgravitating equilibrium matter distributions for varied BH
rotation.  Independent of whether GR effects are included, considering
the same values of temperature and surface area for an isothermal
neutrinosphere, thin disk models yield the highest energy deposition
rates by $\nu\bar\nu$-annihilation, while spherical neutrinospheres
lead to the lowest ones. Considering isothermal neutrinospheres with
the same temperature and surface area, it turns out that compared to
Newtonian calculations, GR effects increase the annihilation rate
measured by an observer at infinity by a factor of 2 when the
neutrinosphere is a disk (in agreement with the previous works;
\citealp{AF01}). However, in case of a torus and a sphere the
influence of GR effects is globally only $\sim 25\%$, although
locally, particularly in the vicinity of the rotation axis of the
system, it can be significantly larger. Focusing on the dependence of
the energy deposition rate on the value of $j_\mathrm{BH}$, it is
found that increasing it from 0 to 1 enhances the energy deposition
rate measured by an observer at infinity by roughly a factor of 2 due
to the change of the inner radius of the neutrinosphere.  Furthermore,
although the absolute values of the energy deposition rate have to be
taken with care (because of the steady state approximation used and
the need of more realistic models for the accretion disk; see
\citealp{BAJM06}, for accretion disks of mass similar to the one
expected in the collapsar model ($M_\mathrm{disk}\simlt 0.01\ms$)
typically, $\dot{E}_{\nu\bar\nu}\sim 10^{50}-10^{51}$\ergsec. Even if
only an small fraction ($\sim 10\%$) of that energy were used to boost
a polar outflow, there is fair chance for neutrinos to be the dominant
energy source of the fireball (at least, in some cases when the
magnetic field is not too large).  The most likely scenario that can
be devised is that both mechanisms (MHD and neutrino energy release)
might be operating simultaneously. Indeed, for MHD-produced jets,
neutrinos will play a fundamental role in the pair-loading of the jet
(e.g., \citealp{LE93}) while, for $\nu$-powered jets, the magnetic
field may be important to collimate the thermally generated outflow.
Which of the two energy deposition mechanisms dominates in every
single GRB will depend on the exact conditions in the precollapse
progenitor.

\section{Outflows emerging from progenitors of short GRBs}
\label{sec:progen_sGRBs}

Nowadays, it is commonly believed that short GRBs (sGRBs) are
generated after the merger of a system compact binaries formed by
either two neutron stars (NSs) or a neutron star and a BH
\citep{Pa86,Go86,Ei89,MH93}. The remnant left by the merger consists
of a newly born BH black hole girded by a thick gas torus from which
it swallows matter at a hypercritical rate. In such situation
radiation is advected inward with the accretion flow and the cooling
is dominated by the emission of neutrinos \citep{PWF99}. As in the
case of progenitors of lGRBs, these neutrinos might either be the
primary energy source blowing a fireball of e$^+$e$^-$ pairs and
photons or to act as mediator in hydromagnetic or electromagnetic
energy extraction mechanisms (see Sect.~\ref{sec:progen_lGRBs}) to
pair-load the Poynting dominated outflow. A fundamental difference
with respect to the collapsar model is that the accretion thick disk
is cannot be continuously refilled from a surrounding matter reservoir
(the stellar matter in case of a collapsar) and, therefore, the
duration of the produced outflows is, in part (see \citealp{AJM05})
roughly limited by the time during needed by the black hole to engulf
most of the matter of the accretion disk, namely, a few 100\,ms. This
limit on the time scale, set by the ON time $t_\mathrm{ce}$ of the
source, of holds for both $\nu$-powered jets and for MHD-generated
outflows. In the first case, the neutrino luminosity fades as the mass
of the disk decreases and, thereby, there will be a critical torus
mass below which a plasma outflow cannot be sustained. In the second
case, for the same reason, there will not be sufficient neutrinos that
pair-load the Poynting dominated jet after a sizable fraction of the
torus has been accreted.
 
Although it seems likely that releasing a few $10^{49}\,$erg above the
poles of a stellar mass BH in a region of nearly vacuum may yield an
ultrarelativistic outflow, numerical simulations are needed to attempt
to answer questions about the collimation mechanism of the polar
outflow, the opening angle of the ultrarelativistic ejecta, the
asymptotic Lorentz factors that can be attained, the internal
structure of the outflow, the duration of a possible GRB event, and
the isotropic equivalent energy which an observer would infer by
assuming the source to expand isotropically. These questions may at
most be guessed, but they cannot be reliably answered on grounds of
merger models and a consideration of their energy release by neutrino
emission and the subsequent conversion of some of this energy by
$\nu\bar\nu$-annihilation to e$^+$e$^Ã¢ÂÂ$-pairs
\citep{RJ99,Jetal99,RR02,RRD03,BAJM06}. Only self-consistently
time-dependent (magneto)hydrodynamic modeling may give us some insight
on the former questions. The reason being that the relativistic outflow
develops in a complex interaction with the accretion torus, cleaning
its own axial funnel such that later energy deposition encounters a
much reduced baryon pollution.

Some keys to answer the questions mentioned in the previous paragraph
have been very recently revealed by time-dependent numerical
simulations in which the main results are:

\noindent 
{\bf Collimation}.  The generated outflows which may yield GRB
signatures are either collimated by the accretion disk \citep{AJM05}
or self-collimated by the magnetic field \citep{McKinney06}, depending
on whether the jet is initiated thermally or magnetically,
respectively. The typical outflow half-opening angles are $\sim 3^o -
25^o$ and, as in the case of jets produced in collapsars, the
baryon-poor outflows display a transverse structure. This structure
shows a central core which spans a half opening angle
$\theta_\mathrm{core}<3^o-12^o$ where $\Gamma_\mathrm{core} > 100$
flanked laterally by a layer, extending up to $\theta_\mathrm{shl}
\sim 25^o$ where the Lorentz factor smoothly decays to moderately
relativistic values and where a sizable fraction of the total energy
is stored. This layer is rather hot in thermally initiated outflows
and has the potential of accelerating even after the energy release by
the central engine has ceased. Similar to the jets produced in
collapsar environments the transverse structure of the Lorentz factor
could be roughly fit by Gaussian profiles. However, somewhat more
complicated functions are required to provide more accurate fits
\citep{AJM05}.

\noindent
{\bf Variability}. Even injecting energy close to the BH even horizon
at constant rates, the produced outflows are highly variable. The
interaction of the newborn fireball with the accretion torus yields
the growth of Kelvin-Helmholtz \citep{AJM05} instabilities.  The
variability in case of MHD jets is imprinted by pinch instabilities
\citep{McKinney06}. Up to date only axisymmetric models have been
computed. All of which seem to be stable or marginally stable. It is
not yet numerically verified whether 3D jets emerging from
hyperaccreting BHs are stable.

\noindent
{\bf Influence of the environment}. Mergers of compact objects may
take usually place in the intergalactic medium or in the outer skirts
of their host galaxies. Thus, the environment of the merger may have a
very low density. However, in the course of the merger, after the
first contact of the two compact objects, there is an ejection of
matter, which is larger close to the orbital plane. Such matter forms
a cool baryon reach environment with a mass $M_\mathrm{halo}<$ a few
$10^{-2}\ms$ \citep{RJ01,OJ06}. The exact amount of mass ejected
sensitively depends on whether the two compact objects are both NSs or
one of them is a BH, on the initial mass ratio between the two,
etc. Thereby, when the central BH forms, the newly born system may, in
some cases, be embedded into a relatively high density halo.  Mergers
in low density environments may fuel ultrarelativistic outflows with
the potential to produce {\it normal} sGRBs, while in case that the
merger occurs in high density media, the observational signature is
not a sGRB but, most likely, a flash in the soft X-ray or UV bands
\citep{AJM05}. In the latter case, the resulting event might only be
observable if it happens very close to the Milky Way. The fact that
depending on the environmental density an sGRB can be produced or not
has the direct implication that not every merger may yield a
sGRB. This fact has to be considered when making estimates about the
true rates of sGRBs and compared with the rates of NS+NS mergers
(e.g., \citealp{GP05}).

\noindent
{\bf Asymptotic Lorentz factor}. The saturation value of the outflow
Lorentz factor, $\Gamma_\infty$, is difficult to estimate on the basis
of numerical simulations that cover only the initial fraction of a
second in the evolution of an ultrarelativistic outflow generated in a
hyperaccreting BH. Despite this difficulties rough estimates can be
made. For instance, in case that the outflows are thermally generated
\citep{AJM05}, there is a clear trend to produce much higher values of
$\Gamma_\infty$ for sGRBs ($\Gamma_\infty \simgt 500 - 1000$) than for
lGRBs ($\Gamma_\infty \sim 100$). The reason for the difference
resides on the much smaller density of the environment of the merger
(even accounting for the mass ejected from the compact objects after
the first contact; see above) as compared with the baryon-polluted
environment that a relativistic jet finds inside a collapsing massive
star. It has been speculated that this difference in Lorentz factor
might be the reason for the paucity of {\em soft} sGRBs
\citep{Jetal06}. The former trend seems not to be followed by
MHD-generated outflows \citep{McKinney06}. The most likely reason
being that McKinney's simulations are set to be scale free, while
there should be a big difference between the environments of mergers
of compact objects and the interior of collapsars. Probably, such a
difference cannot be accommodated easily with simple scale-free
power-laws for the distribution of the physical variables.

  \noindent {\bf Duration of the events}. As pointed out by Aloy et
  al.\@ (2005) and Janka et al.\@ (2006), the "shells" ejected by the
  central engine, accelerate much faster in the leading part of the
  outflow than the shells in its lagging part. The rear shells
  therefore need a longer time to reach velocities $v \simeq c$. This
  differential acceleration at early and late times of the
  relativistic jet leads to a stretching of the overall radial length
  of the outflow, $\Delta$, relative to $t_\mathrm{ce}$ times the
  speed of light $c$, $\Delta > ct_\mathrm{ce}$. This stretching has
  the important consequence that the overall observable duration of
  the GRB (in the source frame), $T = t_\Delta = \Delta/c$, may be a
  factor of 10 or more longer than $t_\mathrm{ce}$, even when the GRB
  is produced by internal shocks.

\section{Summary}

The numerical modelling of progenitors of GRBs has allowed us to gain
some insight into a number of important issues related with the nature
of the outflows produced by these systems. First, it has allowed us to
verify that some (but probably not all) collapsars can yield
collimated relativistic outflows that turn into lGRBs at large
distances from the source. Likewise, only a fraction of the mergers of
compact objects may yield sGRBs. Second, the numerical modeling has
given us information about the collimation mechanism and typical
outflow opening angles. Third, it has shown that the outflow is
heterogeneous both along the direction of propagation and transverse
to it, even if the central engine releases energy at a constant
rate. Fourth, it seems rather plausible that some lGRBs with very high
Lorentz factor ($\Gamma>>100$) need of a MHD jet-formation
mechanism. On the other hand, some other lGRBs with more moderate
Lorentz factors ($\Gamma \sim 100$) can be explained by $\nu$-powered
jets, particularly if $t_\mathrm{ce}<10\,$s.  Fifth, MHD- and
$\nu$-mechanisms may work simultaneously. Therefore, it is likely
that, in some cases MHD processes dominate the jet generation while in
others neutrinos may be the dominant energy source.
	
The numerical modeling done so far is still insuficient. In order to
start from consistent initial models, the state-of-the-art numerical
codes must incorporate, at least, the effects of strong gravity,
magnetic fields and a detailed neutrino transport. In the near future
we may see how all these elements are included in more realistic
numerical experiments that will deepen our understanding on how
ultrarelativistic outflows are produced in progenitors of GRBs.

%\bibstyle{chicago}  

%% \bibliography{general}

\begin{thebibliography}{99}

\bibitem[{Aloy}(2001)]{Aloy01} {Aloy}, M.~A. 2001, in Highlights of
  Spanish Astrophysics II, eds. J.~{Zamorano}, J.~{Gorgas}, and
  J.~{Gallego}, p. 33

\bibitem[{Aloy} et~al.(2000){Aloy}, {M{\"u}ller}, {Ib{\'a}{\~n}ez},
  {Mart{\'{\i}}}, \& {MacFadyen}]{Alo00b}
{Aloy}, M.~A., {M{\"u}ller}, E., {Ib{\'a}{\~n}ez}, J.~M.,
{Mart{\'{\i}}}, J.~M., \& {MacFadyen}, A. 2000, ApJL, 531, L119

\bibitem[{Aloy} et~al.(2002){Aloy}, {Ib{\'a}{\~n}ez}, {Miralles}, \&
  {Urpin}]{Alo02}
{Aloy}, M.~A., {Ib{\'a}{\~n}ez}, J.-M., {Miralles}, J.~A., \& {Urpin}, V.
2002, A\&A, 396, 693

\bibitem[{Aloy} et~al.(2005){Aloy}, {Janka}, \& {M{\"u}ller}]{AJM05}
{Aloy}, M.~A., {Janka}, H.-Th., \& {M{\"u}ller}, E. 2005, ApJ, 436, 273

\bibitem[{Asano} \& {Fukuyama}(2001)]{AF01} {Asano} K., \& {Fukuyama},
  T. 2001, ApJ, 546, 1019

\bibitem[{Birkl} et~al.(2006){Birkl}, {Aloy}, {Janka}, \& {Mueller}]{BAJM06}
{Birkl}, R., {Aloy}, M.~A., {Janka}, H.~-Th., \& {Mueller}, E. 2006, astro-ph/0608543

\bibitem[{Blandford} \& {Znajek}(1977)]{BZ77} {Blandford}, R.~D., \&
  {Znajek}, R.~L. 1977, MNRAS, 179, 433

\bibitem[Blandford \& Payne(1982)]{BP82}
Blandford, R.~D., \& Payne, D.~G. 1982, MNRAS, 199, 883

\bibitem[{Cavallo} \& {Rees}(1978)]{CR78}
{Cavallo} G., \& {Rees}, M.~J. 1978, MNRAS, 183, 359

\bibitem[{De Villiers} et~al.(2005){De Villiers}, {Hawley}, {Krolik}, \&
  {Hirose}]{Villiersetal05}
{De Villiers}, et~al.\@ %J.-P., {Hawley}, J.~F., {Krolik}, J.~H., \& {Hirose}, S.
2005, ApJ, 620, 878

\bibitem[{Eichler} et~al.(1989){Eichler}, {Livio}, {Piran}, \&
  {Schramm}]{Ei89}
{Eichler}, D., {Livio}, M., {Piran}, T., \& {Schramm}, D.~N. 1989,
 Nat., 340, 126

\bibitem[{Frail} et~al.(1997){Frail}, {Kulkarni}, {Nicastro}, {Feroci}, \&
  {Taylor}]{FK97}
{Frail}, D.~A., {Kulkarni}, S.~R., {Nicastro}, S.~R., {Feroci},  M., \& 
  {Taylor}, G.~B. 1997, Nat., 389, 261

\bibitem[{G{\'o}mez} \& {Hardee}(2004)]{GH04} {G{\'o}mez}, E.~A., \&
  {Hardee}, P.~E. 2004, in AIP Conf. Proc. 727: Gamma-Ray Bursts: 30
  Years of Discovery, eds. E.~{Fenimore} \& M.~{Galassi}, p. 278

\bibitem[{Goodman}(1986)]{Go86} {Goodman}, J. 1986, ApJL, 308, L47

\bibitem[{Guetta} \& {Piran}(2005)]{GP05}
{Guetta}, D. \& {Piran}, T. 2005, ApJ, 435, 421

\bibitem[{Harrison} et~al.(1999){Harrison}, {Bloom}, {Frail}, {Sari},
  {Kulkarni}, {Djorgovski}, {Axelrod}, {Mould}, {Schmidt}, {Wieringa}, {Wark},
  {Subrahmanyan}, {McConnell}, {McCarthy}, {Schaefer}, {McMahon}, {Markze},
  {Firth}, {Soffitta}, \& {Amati}]{Harrisonetal99}
 {Harrison}, F.~A.,  et~al.\@ 1999, ApJL, 523, L121

\bibitem[{Heger} et~al.(2005){Heger}, {Woosley}, \& {Spruit}]{HWS05}
{Heger}, A., {Woosley}, S.~E., \& {Spruit}, H.~C. 2005, ApJ, 626, 350

\bibitem[{Janka} et~al.(2006){Janka}, {Aloy}, {Mazzali}, \& {Pian}]{Jetal06}
{Janka}, H.-Th., {Aloy}, M.~A.,  {Mazzali}, P.~A., \& {Pian}, E. 2006, 
ApJ, 645, 1305

\bibitem[{Janka} et~al.(1999){Janka}, {Eberl}, {Ruffert}, \& {Fryer}]{Jetal99}
{Janka}, H.-Th., {Eberl}, Th., {Ruffert}, M., \& {Fryer}, C.~L. 1999, 
ApJ, 527, L39

\bibitem[{Lee} et~al.(2000){Lee}, {Wijers}, \& {Brown}]{LWB00}
{Lee}, H.~K., {Wijers}, R.~A.~M.~J., \& {Brown}, G.~E. 2000, 
Phys. Rep., 325, 83

\bibitem[{Levinson} \& {Eichler}(1993)]{LE93}
{Levinson}, A., \& {Eichler}, D. 1993, ApJ, 418, 386

\bibitem[{Lithwick} \& {Sari}(2001)]{LS01}
{Lithwick}, Y., \& {Sari}, R. 2001, ApJ, 555, 540

\bibitem[{MacFadyen} \& {Woosley}(1999)]{MW99}
{MacFadyen}, A.~I., \& {Woosley}, S.~E. 1999, ApJ, 524, 262

\bibitem[{MacFadyen} et~al.(2001){MacFadyen}, {Woosley}, \& {Heger}]{MWH01}
{MacFadyen}, A.~I.,  {Woosley}, S.~E., \& {Heger}, A. 2001, ApJ, 550, 410

\bibitem[{McKinney}(2005)]{McKinney05}
{McKinney}, J.~C. 2005, ApJL, 630, L5

\bibitem[{McKinney}(2006)]{McKinney06}
{McKinney}, J.~C. 2006,  MNRAS, 368, 1561

\bibitem[{M{\'e}sz{\'a}ros} \& {Rees}(2001)]{MR01}
{M{\'e}sz{\'a}ros}, P., \& {Rees}, M.~J. 2001, ApJL, 556, L37

%\bibitem[{Meszaros} \& {Rees}(1997)]{MR97}
%P.~{Meszaros} \& M.~J. {Rees} 1997, ApJL, 482, L29

\bibitem[{Mizuno} et~al.(2004{\natexlab{a}}){Mizuno}, {Yamada}, {Koide}, \&
  {Shibata}]{Mizuetal04a}
{Mizuno}, Y.~, {Yamada}, S., {Koide}, S., \& {Shibata}, K. 2004a, ApJ, 606, 395

\bibitem[{Mizuno} et~al.(2004{\natexlab{b}}){Mizuno}, {Yamada}, {Koide}, \&
  {Shibata}]{Mizuetal04b}
{Mizuno}, Y.~, {Yamada}, S., {Koide}, S., \& {Shibata}, K. 2004b, ApJ, 615, 389

\bibitem[{Mochkovitch} et~al.(1993){Mochkovitch}, {Hernanz}, {Isern}, \&
  {Martin}]{MH93}
{Mochkovitch}, R., {Hernanz}, M., {Isern}, J., \& {Martin}, X. 1993, Nat., 361, 236

\bibitem[{Obergaulinger} et~al.(2006{\natexlab{a}}){Obergaulinger}, {Aloy},
  {Dimmelmeier}, \& {M{\"u}ller}]{OADM06}
{Obergaulinger}, M., {Aloy}, M.~A., {Dimmelmeier}, H., \&
{M{\"u}ller}, E. 2006a, ApJ, 457, 209

\bibitem[{Obergaulinger} et~al.(2006{\natexlab{b}}){Obergaulinger}, {Aloy}, \&
  {M{\"u}ller}]{OAM06}
{Obergaulinger}, M., {Aloy}, M.~A., \& {M{\"u}ller}, E. 2006b, ApJ, 450, 1107

\bibitem[{Oechslin} \& {Janka}(2006)]{OJ06}
{Oechslin}, R., \& {Janka}, H.-Th. 2006, MNRAS, 368, 1489

\bibitem[{Paczynski}(1986)]{Pa86}
{Paczynski}, B. 1986, ApJL, 308, L43

\bibitem[{Podsiadlowski} et~al.(2004){Podsiadlowski}, {Mazzali}, {Nomoto},
  {Lazzati}, \& {Cappellaro}]{Podsialdlowskietal04}
{Podsiadlowski}, P., et.~al.\@ 2004, ApJL, 607, L17

\bibitem[{Popham} et~al.(1999){Popham}, {Woosley}, \& {Fryer}]{PWF99}
{Popham}, R.,  {Woosley}, S.~E., \& {Fryer}, C. 1999, ApJ, 518, 356

\bibitem[{Ramirez-Ruiz} et~al.(2002){Ramirez-Ruiz}, {Celotti}, \&
  {Rees}]{RRCR02}
{Ramirez-Ruiz}, E., {Celotti}, A., \& {Rees}, M.~J. 2002, MNRAS, 337, 1349

\bibitem[{Rhoads}(1999)]{Rhoads99}
{Rhoads}, J.~E. 1999, ApJ, 525, 737

\bibitem[{Rosswog} \& {Ramirez-Ruiz}(2002)]{RR02}
{Rosswog}, S. \& {Ramirez-Ruiz}, E. 2002, MNRAS, 336, L7

\bibitem[{Rosswog} et~al.(2003){Rosswog}, {Ramirez-Ruiz}, \& {Davies}]{RRD03}
{Rosswog}, S., {Ramirez-Ruiz}, E., \& {Davies}, M.~B. 2003, MNRAS, 345, 1077

\bibitem[{Ruffert} \& {Janka}(2001)]{RJ01}
{Ruffert}, M., \& {Janka}, H.-Th. 2001, ApJ, 380, 544

\bibitem[{Ruffert} \& {Janka}(1999)]{RJ99}
{Ruffert}, M., \& {Janka}, H.-Th, 1999, ApJ, 344, 573

\bibitem[{Sari} et~al.(1999){Sari}, {Piran}, \& {Halpern}]{SPH99}
{Sari}, R., {Piran}, T., \& {Halpern}, J.~P. 1999, ApJL, 519, L17

\bibitem[{Soderberg} \& {Ramirez-Ruiz}(2003)]{SRR03}
{Soderberg}, A.~M., \& {Ramirez-Ruiz}, E. 2003, MNRAS, 345, 854

\bibitem[{Taylor} et~al.(2004){Taylor}, {Frail}, {Berger}, \&
  {Kulkarni}]{TFBK04}
 {Taylor}, G.~B., {Frail}, D.~A., {Berger}, E., \& {Kulkarni}, S.~R. 2004, ApJL, 609, L1

\bibitem[{Woosley}(1993)]{Wo93} {Woosley}, S.~E. 1993, ApJ, 405, 273

\bibitem[{Woosley}(2000)]{Woosley00} {Woosley}, S.~E. 2000, in AIP
  Conf. Proc. 526: Gamma-ray Bursts, 5th Huntsville Symposium, eds.
  R.~M. {Kippen}, R.~S. {Mallozzi}, \& G.~J. {Fishman},  p. 555

\bibitem[{Zhang} et~al.(2003){Zhang}, {Woosley}, \& {MacFadyen}]{ZWM03}
{Zhang}, W., {Woosley}, S.~E., \&  {MacFadyen}, A.~I. 2003, ApJ, 586, 356

\bibitem[{Zhang} et~al.(2004){Zhang}, {Woosley}, \& {Heger}]{ZWH04}
{Zhang}, W., {Woosley}, S.~E., \& {Heger}, A. 2004, ApJ, 608, 365

\end{thebibliography}

\end{document}